\documentclass[amsmath,amssymb,twocolumn,amsfonts]{revtex4}
\usepackage{graphicx}
\usepackage{epsfig}
\usepackage{bm}
\usepackage{bbm,mathbbol}
\usepackage{braket,bigints}
\usepackage{stmaryrd,mathtools}
\def \beq {\begin{equation}}
\def \eeq {\end{equation}}
\def \tr {\rm Tr}

\begin{document}
\title{Quantum trajectories in spin-exchange collisions reveal the nature of spin-noise correlations in multi-species alkali vapors}
\author{K. Mouloudakis$^1$}
\author{M. Loulakis$^{2,3}$}
\author{I. K. Kominis$^{1,4}$}
\email{ikominis@physics.uoc.gr}
\affiliation{$^1$Department of Physics, University of Crete, Heraklion 71003, Greece\\
$^2$School of Applied Mathematical and Physical Sciences, National Technical University of Athens, 15780 Athens, Greece\\
$^3$Institute of Applied and Computational Mathematics, FORTH, 70013 Heraklion, Greece\\
$^4$Institute of Theoretical and Computational Physics, University of Crete, Heraklion 70013, Greece}

\begin{abstract}
Spin-exchange collisions in alkali vapors underly several fundamental and applied investigations, like nuclear structure studies and tests of fundamental symmetries, ultra-sensitive atomic magnetometers, magnetic resonance and bio-magnetic imaging. Spin-exchange collisions cause loss of spin coherence, and concomittantly produce spin noise, both phenomena being central to quantum metrology. We here develop the quantum trajectory picture of spin-exchange collisions, consistent with their long-standing ensemble description using density matrices. We then use quantum trajectories to reveal the nature of spin-noise correlations that spontaneously build up in multi-species atomic vapors, frequently utilized in the most sensitive spin measurements.
\end{abstract}
\maketitle 
\section{Introduction}
Atomic spin-exchange collisions are fundamental for a broad range of explorations, from nuclear physics and astrophysics to atomic spectroscopy, quantum metrology and medical imaging. Spin-exchange collisions have been studied in the context of hyperfine transitions in hydrogen masers \cite{Walsworth_1990} and radio emission of interstellar hydrogen \cite{Zygelman_2005}. Spin-exchange collisions in alkali vapors have been central in producing non-equilibrium magnetic substate populations by optical pumping \cite{Happer_RMP_1972}. Spin-exchange optical pumping \cite{Walker_Happer_RMP_1997,Appelt_PRA_1998} has led to hyper-polarized noble gases used in medical imaging \cite{Albert}, spectroscopy \cite{Pines} and numerous studies of nucleon spin structure \cite{Deur}. Furthermore, the intricate physics of spin-exchange collisions \cite{Happer_Tam_PRA_1977} have spurred the development of ultra-sensitive magnetometers \cite{Budker_Romalis}, allowing new precision tests of fundamental symmetries \cite{Romalis_2011,BudkerRMP} and novel biomagnetic imaging applications \cite{Xia,Kitching}. Spin-exchange collisions cause spin coherence relaxation, and since relaxation and fluctuations are intimately connected, spin-exchange collisions are also underpinning spin noise spectroscopy \cite{sn1,sn2,sn3,sn4,sn5,sn_review,Mitchell_2016,Mitchell}. 

So far, however, the understanding of spin-exchange collisions was based on ensemble descriptions with density matrix master equations, which neither capture the physics at the single-atom level, nor the spontaneous fluctuations of the collective spin or any phenomena stemming therefrom. We here use quantum measurement theory to shed light on the quantum foundations of spin-exchange collisions. We develop the quantum-trajectory picture of spin-exchange collisions \cite{Deutsch}, and demonstrate the consistency of our unravelling with the well-established density matrix master equation \cite{Happer_Book}. Quantum trajectories can be used to understand at the single-atom level alkali-alkali, or even alkali-nobel gas collisions \cite{Firstenberg}. We then demonstrate how quantum trajectories can seamlessly produce spin noise from first principles. Moving to dual-species vapors, we use quantum trajectories to reveal the nature of spin-noise correlations that spontaneously build up \cite{Dellis,Roy}. As a byproduct of the above, we present a first approach to the stochastic terms augmenting the density matrix master equation describing spin-exchange collisions.

The structure of this article is the following. In Section II we briefly reiterate the long-standing ensemble description of spin-exchange collisions in hot atomic vapors, which uses a density matric master equation. In Section III we introduce the quantum-trajectory picture of spin-exchange collisions, and demonstrate its consistency with the ensemble description is physical scenarios involving large non-equilibrium spin polarizations but not fluctuations, i.e. in scenarios tractable with the master equation. We then show how quantum trajectories can seamlessly produce spin noise, which cannot be accounted for by the spin-exchange master equation. In Section IV we move to dual-species vapors, and establish the consistency of the quantum-trajectory approach with the coupled master equations of the ensemble approach, again in physical scenarios involving large non-equilibrium spin polarizations but not fluctuations. Finally, in Section V we use quantum trajectories to address an open problem in spin-noise spectroscopy, namely the spin-noise correlations that spontaneously build up in dual-species vapors. Finally, we augment the master equations describing spin-exchange collisions with physically realistic noise terms, rendering the description of spin noise possible at the ensemble level. The stochastic master equations provide an independent verification of the nature of spin-noise correlations. 
\section{Spin-exchange collisions: ensemble description}
Spin-exchange (SE) collisions between two atoms A and B, of the same or different species, result from the different interaction potentials, $V_S$ and $V_T$, for the singlet and triplet total spin of the colliding partners, respectively. If $\mathbf{s}_a$ and $\mathbf{s}_b$ are the electron spins of the colliding atoms, the singlet and triplet projectors are ${\cal P}_S={1\over 4}\mathbb{1}-\mathbf{s}_a\cdot\mathbf{s}_b$ and ${\cal P}_T={3\over 4}\mathbb{1}+\mathbf{s}_a\cdot\mathbf{s}_b$ \cite{note1}. Introducing the exchange operator ${\cal P}_e={\cal P}_T-{\cal P}_S$, the SE interaction potential $V_{\rm se}={\cal P}_{S}V_S+{\cal P}_TV_T$ is written as a sum of a spin-independent and a spin-dependent term, $V_{\rm se}=V_0\mathbb{1}+V_1{\cal P}_e$. Only the latter is of interest for the unitary operator evolving the initial into the final spin state, $U=e^{-i\int dt V_1{\cal P}_e}$ \cite{note11}. Defining $\phi=\int dt V_1$
and noting that ${\cal P}_e^2=\mathbb{1}$, we get $U=\cos\phi\mathbb{1}-i\sin\phi {\cal P}_e$ \cite{Brun_AJP_2002}.

For a single-species vapor of atoms A with number density [A] the ensemble description of SE collisions follows either \cite{Grosettete} by using ${\cal P}_e$ and considering the SE rate $1/T_{\rm se}=[{\rm A}]\overline{v}\sigma_{\rm se}$, or by using $U$  \cite{Happer_Book} and identifying $\overline{\sin^2\phi}/T$ with $1/T_{\rm se}$, where $T$ the time between collisions and $\overline{\sin^2\phi}$ the collisional average of $\sin^2\phi$. Here $\sigma_{\rm se}$ is the SE cross section and $\overline{v}$ the mean relative velocity of the colliding atoms. Neglecting the SE frequency shift \cite{Happer_Book}, both approaches result in the same master equation. For following use, we here briefly reiterate the first approach.

Let two atoms A and B collide, initially assuming they are of different species, with their (uncorrelated) pre-collision states being $\rho_a$ and $\rho_b$. Hence the combined two-atom initial state is $\rho_a\otimes\rho_b$.
The state of atom A after the collision is $\rho_{ab}^e=\tr_{B}\{{\cal P}_e\rho_a\otimes\rho_b{\cal P}_e\}$,
\begin{align}
\rho_{ab}^e&={1\over 4}\rho_a+\mathbf{s}_a\cdot\rho_a\mathbf{s}_a+(\rho_a\mathbf{s}_a+\mathbf{s}_a\rho_a)\cdot\braket{\mathbf{s}_b}\nonumber\\
&-2i\braket{\mathbf{s}_b}\cdot(\mathbf{s}_a\times\rho_a\mathbf{s}_a)\label{rhoABe}
\end{align}
For treating a single-species vapor, one substitutes $b\rightarrow a$ and arrives at the master equation describing both Hamiltonian evolution and SE collisions:
\beq
d\rho_a/dt=-i[{\cal H}_a,\rho_a]-(\rho_a-\rho_{aa}^e)/T_{\rm se}\label{ME},
\eeq
where ${\cal H}_a=\omega s_{az} +A\mathbf{s}_a\cdot\mathbf{I}_a$ is the alkali ground-state Hamiltonian in the presence of a $z$-axis magnetic field $\omega$, and $A$ the hyperfine coupling with the nuclear spin $\mathbf{I}_a$. The second term in Eq. \eqref{ME} is responsible for transverse spin relaxation \cite{Savukov}. The above are well-known results comprehensively presented in \cite{Happer_Book}.
\begin{figure}[t!]
\begin{center}
\includegraphics[width=8 cm]{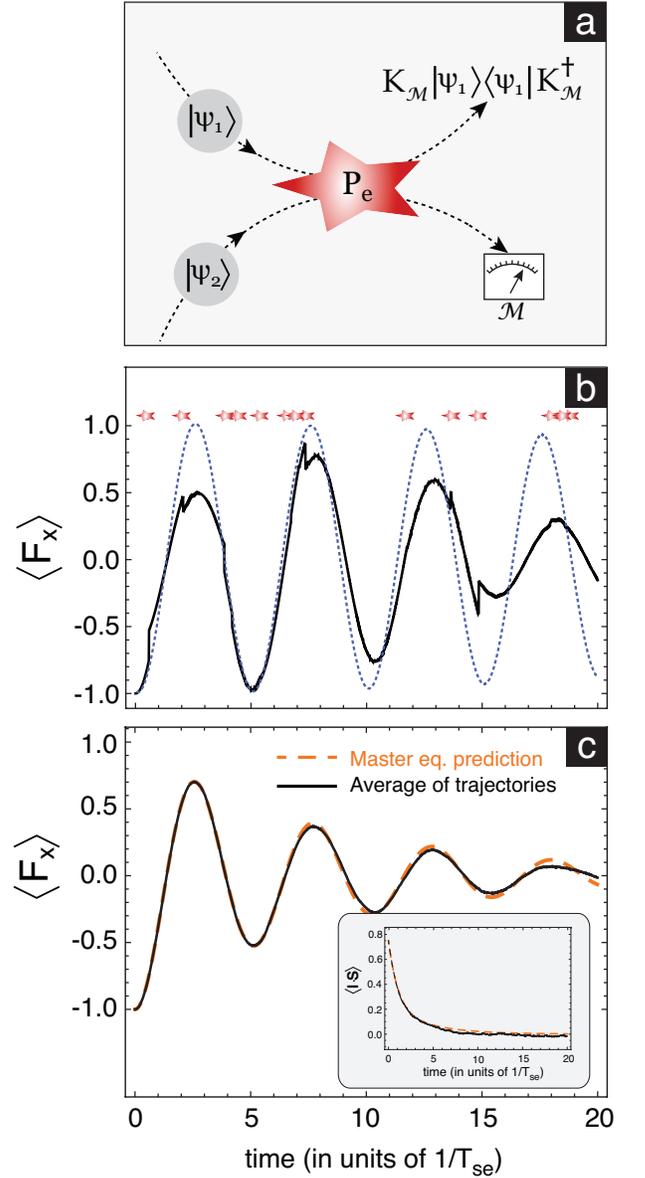}
\caption{(a) Quantum measurement picture of a binary spin-exchange collision. (b) Single-atom trajectory expectation value of $\braket{F_x}$. Red stars depict the random occurrences of SE collisions. Chosen parameters were $I=3/2$, and in units of $1/T_{\rm se}$, $\omega=5$ and $A=100$, while time from $t=0$ to $t=20$ was split into 6M steps. Initial state was the $\ket{2-1}$ eigenstate of $F_x$. Blue dotted line is just the Hamiltonian evolution. (c) Average (black solid line) of 4000 trajectories like in (b), compared with the predictions of Eq. \eqref{ME} (orange dashed line). (inset) similar but for the observable $\braket{\mathbf{I}\cdot\mathbf{s}}$.}
\label{fig1}
\end{center}
\end{figure}
\section{Spin-exchange collisions: quantum-trajectory description}
We will now develop the quantum trajectory approach, pictured in Fig. \ref{fig1}a, that is consistent with the above ensemble description. We start with two, initially uncorrelated, atoms 1 and 2 being in the pure states $\ket{\psi_1}$ and $\ket{\psi_2}$. The combined two-atom pre-collision and post-collision state is $\ket{\psi_1}\otimes\ket{\psi_2}$ and ${\cal P}_e\ket{\psi_1}\otimes\ket{\psi_2}$, respectively. 
We now consider atom 1 as the  "system", and atom 2 as the "probe", which we can extract information about the system. To do so, we perform a projective measurement on the probe. For this we here use the $\ket{FM}$ basis (zero-field eigenstates of $\mathbf{I}\cdot\mathbf{s}$ and $F_z$), but any other complete basis would do equally well. Atom 2 is projected in some state $\ket{FM}$ with probability given by the norm of the state $\Pi_{FM}{\cal P}_e\ket{\psi_1}\otimes\ket{\psi_2}$, where $\Pi_{FM}=\mathbb{1}\otimes\ket{FM}\bra{FM}$. Defining the Kraus operator corresponding to the particular measurement outcome on atom 2 as ${\cal K}_{FM}=\bra{FM}{\cal P}_e\ket{\psi_2}$, we find that $p_{FM}=\bra{\psi_1}{\cal K}_{FM}^{\dagger}{\cal K}_{FM}\ket{\psi_1}$.
Concomitantly, atom 1 is projected to 
\beq
\ket{\psi_1^e}_{FM}={{{\cal K}_{FM}\ket{\psi_1}}\over \sqrt{p_{FM}}}\label{psiSE}
\eeq
The Kraus operators satisfy the completeness relation $\sum_{FM}{\cal K}^{\dagger}_{FM}{\cal K}_{FM}=\mathbb{1}$, which readily follows from the completeness of the $\ket{FM}$ basis states and the property ${\cal P}_e^2=\mathbb{1}$ \cite{note2}. Hence it is also $\sum_{FM}p_{FM}=1$, as it should be. Finally, the state $\ket{\psi_1^e}_{FM}$ is properly normalized. 

For the numerical production of quantum trajectories, we consider $N$ atoms, each in any desired pure initial state. We split time into intervals $dt$ \cite{note3}, and in every $dt$ we evolve each atom unitarily with the Hamiltonian. Moreover, in each $dt$ there is a probability, $dt/T_{\rm se}$, that an atom suffers an SE collision. A random number drawn for each atom decides whether this probability is realized. For those atoms ("system" atom 1) undergoing an SE collision during $dt$, the collision partner ("probe" atom 2) is randomly chosen from the same list of $N$ atoms.  Finally, given $p_{FM}$, we let another random number choose what is the projective measurement outcome $\ket{FM}$ of atom 2. Then atom 1 is projected to $\ket{\psi_1^e}_{FM}$ given by Eq.\eqref{psiSE}. 

Crucially, although the particular projection of the "probe" atom to a state $\ket{FM}$ determines the projection of the "system" atom to $\ket{\psi_1^e}_{FM}$, after the collision we leave atom 2 in its initial pre-collision state. This has imprinted a Markovian interpretation, i.e. all "probe" atoms instantly loose memory of the collisions taking place during $dt$ and re-constitute the ensemble pre-collision state. We here do not investigate whether the above picture is physically precise, but only care to lay out the single-atom physics behind the master equation \eqref{ME}, which has been impressively successful in accounting for a broad set of experimental data.

Having access to $N$ quantum trajectories, we can evaluate the time-evolution of any observable $Q$ either for the particular $j$-th trajectory, as $\braket{Q}^{(j)}_t=\braket{\psi_j(t)|Q|\psi_j(t)}$, or for the trajectory-average used to compare with the master equation, as $\braket{Q}_t={1\over N}\sum_{j=1}^{N}\braket{Q}^{(j)}_t$. In Fig. \ref{fig1}b we choose for $Q$ the transverse spin $F_x$ and show an example of a single trajectory depicting the discontinuities due to SE collisions. In Fig. \ref{fig1}c we show how the average of many such trajectories accounts for spin relaxation at the ensemble level, and compare the trajectory average with Eq. \eqref{ME}, both for $F_x$ and for a second observable, $\mathbf{I}\cdot\mathbf{s}$. The perfect agreement demonstrates the consistency of our quantum trajectory approach.
\begin{figure}[t!]
\begin{center}
\includegraphics[width=8 cm]{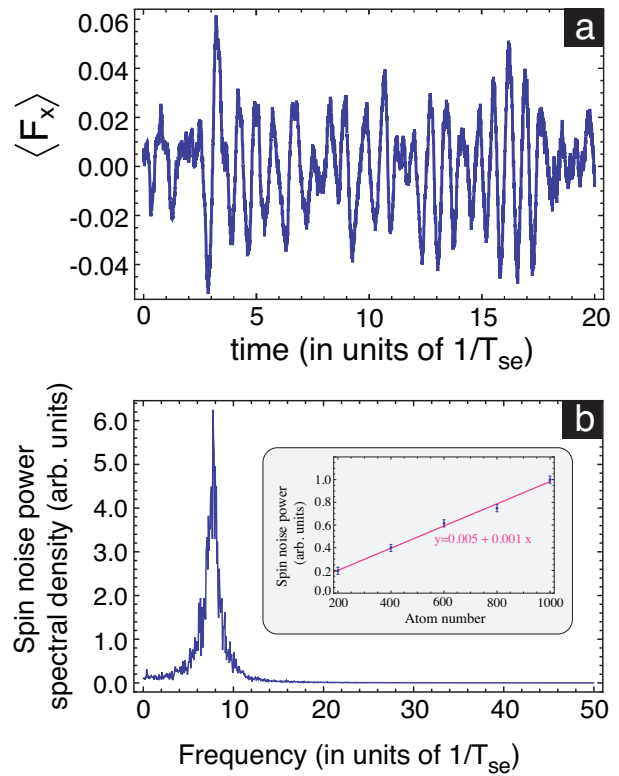}
\caption{(a) Spin noise produced by randomly chosing one of the eigenstates of $F_x$ as the initial state for each of 400 trajectories. Here $I=3/2$, $\omega=32$ and $A=100$. (b) Spin noise spectrum resulting from the average of 50 FFT power spectra derived from 50 time traces like (b). Inset shows the linear dependence of the total spin noise power of a single FFT power spectrum as a function of atom number. Error bars were estimated from two different FFT power spectra.}
\label{fig2}
\end{center}
\end{figure}

We next show how the trajectory picture produces spin noise. We chose for each of the $N$ atoms a random initial state among all eigenstates of $F_x$. In Fig. \ref{fig2}a we plot the resulting fluctuations of $\braket{F_x}$, and in Fig. \ref{fig2}b an FFT spectrum that is typically observed in spin-noise spectroscopy. We note that the spin noise in Fig. \ref{fig2}a is not due to an imbalance of the initial states in a finite number of $N$ trajectories. Similar spin noise traces can be produced even by exactly balancing the initial states so that at $t=0$ it is $\braket{F_x}=0$, and even by using as initial states eigenstates of $F_z$. Hence spin noise produced by SE collisions is a genuine quantum effect originating from the quantum randomness of the post-collision states.
\section{Spin-exchange collisions in dual-species vapors: Ensemble versus quantum trajectory approach}
We now move to the case of a dual-species vapor of atoms A and B with relative abundances $\eta_a$ and $\eta_b$ ($\eta_a+\eta_b=1$). This is treated similarly, using $N$ atoms with ground-state Hamiltonian $H_a$ and another $N$ atoms with ground-state Hamiltonian $H_b$. Now we need to introduce four kinds of collisions, i.e. A-A, A-B, B-A and B-B collisions, with the respective relaxations rates, $\gamma_{\alpha\beta}$, and SE collision probabilities, $dP_{\alpha\beta}=\gamma_{\alpha\beta}dt$. The rates $\gamma_{\alpha\beta}$ are proportional to the mean relative $\alpha-\beta$ velocity and to the abundance $\eta_\beta$, i.e. $\gamma_{\alpha\beta}\propto\overline{v}_{\alpha\beta}\eta_\beta$, since the SE relaxation rate of the first collision partner is proportional to the atom number density of the second collision partner. Given $dP_{{\alpha\beta}}$, we let random numbers decide whether each among the $2N$ atoms will perform a "self"- and/or a "cross"-exchange collision, and we again randomly choose the collision partner. 
\begin{figure}[t!]
\begin{center}
\includegraphics[width=7.6 cm]{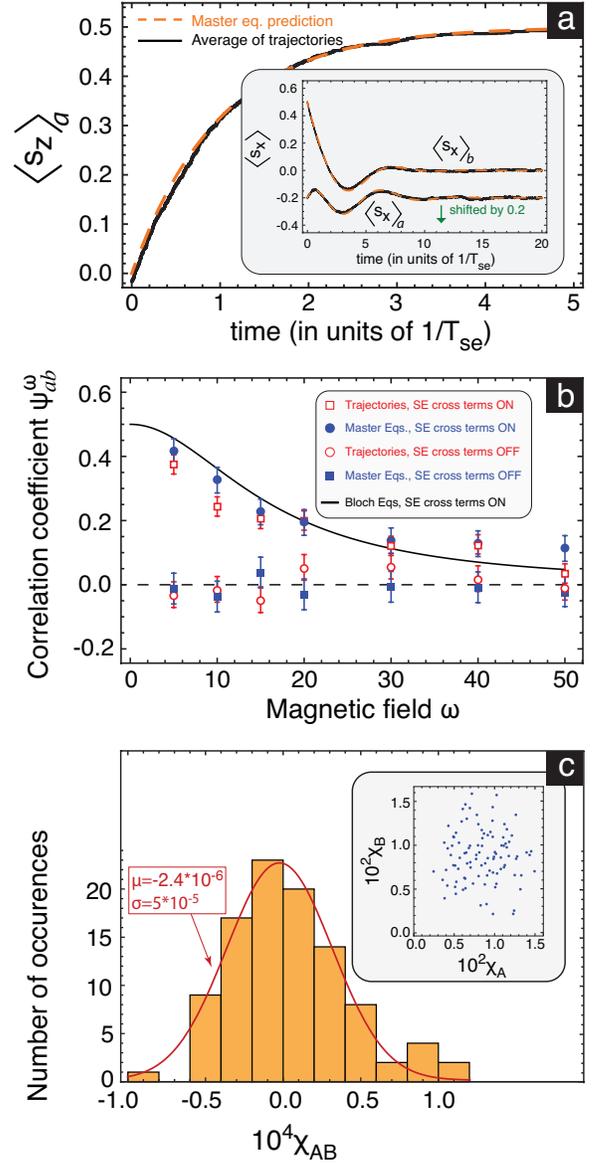}
\caption{(a) Spin transfer from 1000 atoms B ($I=5/2$), always kept in a $+1/2$ eigenstate of $s_z$, to 1000 atoms A ($I=3/2$), initially unpolarized. Trajectory average (black solid line), compared with Eq. \eqref{cME1} keeping only the cross-exchange term (dashed orange line). (inset) Similar comparison with the B atoms initially in the $\ket{33}$ eigenstate of $F_x$. For both vapors we took $A=50$ and $\omega=5$. The trace for $\braket{s_x}_a$ is shifted downwards for clarity. (b) Spin-noise correlation coefficient $\psi_{ab}^\omega$ for 400 atoms of type A and 400 atoms of type B. Time from $t=0$ to $t=20$ was split into 2M steps. Initially, each atom is in a random eigenstate of $F_{x}$. Every point is the mean of 10 runs, and the error bar the standard error of the mean. Shown are the trajectory average and the stochastic coupled master equations prediction, with and without cross-exchange. Solid line is the prediction of the theory developed in \cite{Dellis}. (c) Distribution of $\chi_{ab}$, $\chi_a$ and $\chi_b$ for 100 spin noise runs with randomized $5\leq\omega\leq 50$ and $0.1\leq\gamma_{\alpha\beta}\leq 1.5$.}
\label{fig3}
\end{center}
\end{figure}

Extending Eq. \eqref{ME} to a dual-species vapor we arrive at the two coupled master equations (derived in \cite{Happer_Book})
\begin{align}
{{d\rho_a}\over {dt}}&=-i[{\cal H}_a,\rho_a]-\gamma_{aa}(\rho_a-\rho_{aa}^e)-\gamma_{ab}(\rho_a-\rho_{ab}^e)\label{cME1}\\
{{d\rho_b}\over {dt}}&=-i[{\cal H}_b,\rho_b]-\gamma_{ba}(\rho_b-\rho_{ba}^e)-\gamma_{bb}(\rho_b-\rho_{bb}^e)\label{cME2},
\end{align}
where $\rho_{ba}^e$ and $\rho_{bb}^e$ follow from $\rho_{ab}^e$ in Eq. \eqref{rhoABe} by exchanging $a\leftrightarrow b$, and by substituting $a\rightarrow b$, respectively.

We provide two examples of the consistency between Eqs. \eqref{cME1},\eqref{cME2} and the trajectory average. In Fig. \ref{fig3}a we demonstrate the transfer of spin polarization from type-B atoms to type-A atoms. This is described by the simple equation $d\braket{s_z}_a/dt=(\braket{s_z}_b-\braket{s_z}_a)/T_{\rm se}$ \cite{HapperPrimer}, and can be experimentally realized by optically pumping with circularly polarized light atoms B, and measuring the spin polarization building up in atoms A. To translate these dynamics into the trajectory picture we use a $+1/2$ eigenstate of $s_z$ as the initial state of atoms B, and an equal mixture of eigenstates of $s_z$ for atoms A. Atoms B are always kept in their initial state, and we consider only A-B collisions. Next we consider coherent dynamics, by setting the initial spin polarization of atoms B transversely to the magnetic field, and by again having atoms A unpolarized. We now observe simultanesouly (i) the precession, (ii) the cross-transfer, and (iii) the decay of the transverse polarizations, as shown in the inset of Fig. \ref{fig3}a. Perfect agreement is again observed for both cases.
\section{Spin-noise correlations in dual species vapors}
Having established the consistency of our trajectory approach with the coupled dynamics of Eqs. \eqref{cME1} and \eqref{cME2}, we now move to analyze spin noise correlations that spontaneously build up in coupled vapors \cite{Dellis, Roy}. Like in Fig. \ref{fig2}a, we generate spin-noise time traces $\braket{s_x}_a^{(j)}$ and $\braket{s_x}_b^{(j)}$ in a dual-species vapor, and do so for various magnetic fields $\omega$. The index $j$ runs from $j=1,...,j_{\rm max}$, where $j_{\max}dt$ is the considered total time interval. In Fig. \ref{fig3}b we show the correlation coefficient $\psi_{ab}^\omega$ for each magnetic field value $\omega$, where
\beq
\psi_{ab}^\omega={{\sum_{j}(\braket{s_x}_a^{(j)}-\overline{\braket{s_x}}_a)(\braket{s_x}_b^{(j)}-\overline{\braket{s_x}}_b)}\over\sqrt{\sum_{j}(\braket{s_x}_a^{(j)}-\overline{\braket{s_x}}_a)^2\sum_{j}(\braket{s_x}_b^{(j)}-\overline{\braket{s_x}}_b)^2}}
\eeq
At low $\omega$ we observe positive correlations, which then tend to zero as $\omega$ increases. This effect was measured in \cite{Dellis}, and was theoretically derived from the coupled Bloch equations augmented with noise generating terms. Here the positive correlation effect is demonstrated with a first-principles quantum trajectory analysis without any assumption. The coupled Bloch equations prediction is also shown in Fig. \ref{fig3}c for completeness. 

Importantly, we here move beyond the Bloch equations, and further support the positive correlations effect using Eqs. \eqref{cME1} and \eqref{cME2}. To do so, these equations need to be augmented with noise terms, which are operators acting in the relevant Hilbert space, whereas in the Bloch equations \cite{Dellis} the noise terms were just c-numbers. The first-principles derivation of these noise operators will be addressed elsewhere. Here we make an ad-hoc, but physically realistic assumption about their form, and show that the stochastic master equations prediction for $\psi_{ab}^\omega$ agrees with the trajectory average. 

Explicitly, since spin noise produces spin polarization of order $1/\sqrt{N}\ll1$, we assume that the atom's density matrix describing spin noise produced by SE collisions follows a spin-temperature distribution \cite{Appelt_PRA_1998} with a fluctuating albeit high temperature, i.e. $\rho=e^{-\beta F_x}/\tr\{e^{-\beta F_x}\}$, with $\beta\ll 1$. Expanding $\rho$ around $\beta=0$ we find for the differential change $\delta\rho\propto F_x$. Hence we set for the stochastic terms added to \eqref{cME1} and \eqref{cME2} $\delta_t\rho_a=\sqrt{\gamma_a\over N_a}F_xd\xi^a_t$ and $\delta_t\rho_b=\sqrt{\gamma_b\over N_b}F_xd\xi^b_t$ \cite{note4}, respectively, where $\gamma_a=\gamma_{aa}+\gamma_{ab}$ and $\gamma_b=\gamma_{ba}+\gamma_{bb}$ \cite{note5}, while $d\xi^a_t$ and $d\xi^b_t$ are real and {\it independent} Wiener processes with zero mean and variance $dt$, i.e. $\langle\langle d\xi^a_t\rangle\rangle=\langle\langle d\xi^b_t\rangle\rangle=0$ and $\langle\langle d\xi^\alpha_td\xi^\beta_{t'}\rangle\rangle=dt\delta_{\alpha\beta}\delta(t-t')$, with $\alpha,\beta=a,b$. The prediction of the stochastic master equations is shown in Fig. \ref{fig3}b to reproduce the trajectory average. 

As a systematic check, we turn off the cross-exchange process in the generation of trajectories, i.e. we do not perform A-B and B-A collisions. Similarly, we turn off the cross-exchange coupling terms in the stochastic coupled master equations. As shown in Fig. \ref{fig3}b, $\psi_{ab}^\omega$ is consistent with zero for both cases. 

Finally, the fact that the noise terms $d\xi^a_t$ and $d\xi^b_t$ should be independent can be further supported by the quantum trajectories, from which we 
calculate $\chi_{aa}$, $\chi_{bb}$ and $\chi_{ab}$, where $\chi_{\alpha\beta}=\sum_j(\braket{s_x}_\alpha^{(j+1)}-\braket{s_x}_\alpha^{(j)})(\braket{s_x}_\beta^{(j+1)}-\braket{s_x}_\beta^{(j)})$, with $\alpha,\beta=a,b$. We do this for 100 spin noise runs, with randomized rates $\gamma_{\alpha\beta}$ and magnetic field $\omega$. In Fig. \ref{fig3}c it is seen that $\chi_{ab}$ is about three orders of magnitude less than $\chi_a$ and $\chi_b$. Yet, a standard result \cite{Revuz} on the quadratic variation of an Ornstein-Uhlenbeck diffusion process is that (as $dt\rightarrow 0$) $\chi_{\alpha\beta}\propto\langle\langle d\xi^\alpha_td\xi^\beta_{t}\rangle\rangle$. Thus, spin-noise correlations in dual species vapors are consistent with independent noise terms driving the master equations \eqref{cME1} and \eqref{cME2}, the correlations being produced by the cross-couplings terms in \eqref{cME1} and \eqref{cME2} and not by any conspicuous choice of the noise terms.
\section{Conclusions}
In conclusion, we have developed the single-atom quantum trajectory picture of spin-exchange collisions consistent with the density matrix ensemble descirption used so far. This picture is ideally suitable to understand quantum fluctuations of all sorts of spin observables, the fluctuations being driven by the incessant atomic collisions and the resulting binary spin exchange interaction. As a first application, we demonstrated from first principles that spin-exchange collisions spontaneously produce positive spin-noise correlations in vapors containing two atomic species. 
\acknowledgements
K.M. acknowledges the co-financing of this research by Greece and the European Union (European Social Fund- ESF) through the Operational Programme «Human Resources Development, Education and Lifelong Learning» in the context of the project "Strengthening Human Resources Research Potential via Doctorate Research” (MIS-5000432), implemented by the State Scholarships Foundation (IKY)".


\begin{thebibliography}{0}

\bibitem{Walsworth_1990}
R. L. Walsworth, I. F. Silvera, E. M. Mattison and R. F. C. Vessot, Phys. Rev. Lett. {\bf 64}, 2599 (1990).

\bibitem{Zygelman_2005}
B. Zygelman, Astrophys. J. {\bf 622}, 1356 (2005).

\bibitem{Happer_RMP_1972}
W. Happer, Rev. Mod. Phys. {\bf 44}, 169 (1972).

\bibitem{Walker_Happer_RMP_1997}
T. G. Walker and W. Happer, Rev. Mod. Phys. {\bf 69}, 629 (1997).

\bibitem{Appelt_PRA_1998}
S. Appelt, A. B.-A. Baranga, C. J. Erickson, M. V. Romalis, A. R. Young and W. Happer, Phys. Rev. A {\bf 58}, 1412 (1998).

\bibitem{Albert}
M. S. Albert, G. D. Cates, B. Driehuys, W. Happer, B. Saam, C. S. Springer, Jr., and A. Wishnia, Nature (London) {\bf 370}, 199 (1994).

\bibitem{Pines}
G. Navon, Y. Q. Song, T. Room, S. Appelt, R. E. Taylor and A. Pines, Science {\bf 271}, 1848 (1996).

\bibitem{Deur}
A. Deur, S. J. Brodsky and G. F. de T\'{e}ramond, Rep. Prog. Phys. {\bf 82}, 076201 (2019).

\bibitem{Happer_Tam_PRA_1977}
W. Happer and A. C. Tam, Phys. Rev. A {\bf 16}, 1877 (1977).

\bibitem{Budker_Romalis}
D. Budker and M. V. Romalis, Nature Physics {\bf 3}, 227 (2007).

\bibitem{Romalis_2011}
M. Smiciklas, J. M. Brown, L. W. Cheuk, S. J. Smullin and M. V. Romalis, Phys. Rev. Lett. {\bf 107}, 171604 (2011).

\bibitem{BudkerRMP}
M. S. Safronova, D. Budker, D. DeMille, Derek F. Jackson Kimball, A. Derevianko and Charles W. Clark, Rev. Mod. Phys. {\bf 90}, 025008 (2018).

\bibitem{Xia}
H. Xia, A. Ben-Amar Baranga, D. Hoffman and M. V. Romalis, Appl. Phys. Lett. {\bf 89}, 211104 (2006).

\bibitem{Kitching}
T. Sander, J. Preusser, R. Mhaskar, J. Kitching, L. Trahms and S. Knappe, Biomed. Opt. Express {\bf 3}, 981 (2012).

\bibitem{sn1}
E. B. Aleksandrov and V. S. Zapasskii,  Sov. Phys. JETP {\bf 54} 64-67 (1981).

\bibitem{sn2}
S. A. Crooker, D. G. Rickel, V. A. Balatsky and D. L. Smith, Nature {\bf 431}, 49-52 (2004).

\bibitem{sn3}
M. Oestreich, M. R\"{o}mer, R. J. Haug and D. H\"{a}gele, Phys. Rev. Lett. {\bf 95}, 216603 (2005).

\bibitem{sn4}
G. E. Katsoprinakis, A. T.  Dellis and I. K. Kominis, Phys. Rev. A {\bf 75}, 042502 (2007).

\bibitem{sn5}
V. S. Zapasskii {\it et al.}, Phys. Rev. Lett. {\bf 110}, 176601 (2013).

\bibitem{sn_review}
N. A. Sinitsyn and Y. V. Pershin, Rep. Prog. Phys. {\bf 79} 106501 (2016).

\bibitem{Mitchell_2016}
V. G, Lucivero, R. Jim\'{e}nez-Mart\'{i}nez, J. Kong and M. W. Mitchell, Phys. Rev. A {\bf 93}, 053802 (2016).

\bibitem{Mitchell}
J. Kong, R. Jim\'{e}nez-Mart\'{i}nez, C. Troullinou, V. G. Lucivero and M. W. Mitchell, arXiv:1804.07818.

\bibitem{Deutsch}
I. H. Deutsch and P. S. Jessen, Opt. Commun. {\bf 283}, 681 (2010).

\bibitem{Happer_Book}
W. Happer, Y.-Y. Jau and T. G. Walker, {\it Optically pumped atoms}, Wiley-Vch Verlag Gmbh \& Co. KGaA, Weinheim, 2010.

\bibitem{Firstenberg}
O. Katz, R. Shaham and O. Firstenberg, arXiv:1905.12532.

\bibitem{Dellis}
A. T. Dellis, M. Loulakis and I. K. Kominis, Phys. Rev. A {\bf 90}, 032705 (2014).

\bibitem{Roy}
D. Roy, L. Yang, S. A. Crooker and N. A. Sinitsyn, Sci. Rep. {\bf 5}, 9573 (2015).

\bibitem{note1}
We use $\mathbb{1}$ to denote the unit matrix of dimension that follows from the context, e.g. for two-body operators like ${\cal P}_e$, the unit matrix refers to the combined two-atom Hilbert space of dimension $2(2I_A+1)2(2I_B+1)$.

\bibitem{note11}
The time dependence of $V_1$ is implicit in its dependence on the internuclear distance, which changes with time along the collision trajectory.

\bibitem{Brun_AJP_2002}
In the quantum information literature ${\cal P}_e$ and $U$ are more widely known as the SWAP and partial-SWAP operator, respectively, see e.g. T. A. Brun, Am. J. Phys. {\bf 70}, 719 (2002).

\bibitem{Grosettete}
F. Grosett\^{e}te, J. Phys. (Paris) {\bf 25}, 283 (1965); {\bf 29}, 456 (1968).

\bibitem{Savukov}
I. M. Savukov and M. V. Romalis, Phys. Rev. A {\bf 71}, 023405 (2005).

\bibitem{note2}
Here the unit operator refers to the single-atom Hilbert space, because ${\cal K}_{FM}$ are single-atom operators.

\bibitem{note3}
The value of $dt$ is chosen so that oscillating observables are numerically well-represented. Since realistic values of $\omega$ and $A$ in the Hamiltonian ${\cal H}$ would require such a small value of $dt$, that the simulation would not run in realistic times, we here chose "unrealistically" small hyperfine couplings, which in any case do not affect the physical considerations herein.

\bibitem{HapperPrimer}
W. Happer, Hyperfine. Int. {\bf 38}, 35 (1987).

\bibitem{note4}
We note that the operator $F_x$ is a different matrix in the two noise operators, since they refer to different atom species (i.e. with different nuclear spin), and hence different Hilbert spaces.

\bibitem{note5}
The particular numerical value in front of the operator $F_x$ in the noise terms is inconsequential for this work, since it drops out of the A-B correlation coefficient.

\bibitem{Revuz}
D. Revuz and M. Yor, {\it Continuous Martingales and Brownian Motion} (Springer Science \& Business Media, 2013).

\end{thebibliography}
\end{document}